# Experiments on Data Preprocessing of Persian Blog Networks

Zeinab Borhani-Fard
School of Computer Engineering
University of Qom
Qom, Iran

Leila Esmaeili
School of Computer Science & Information Technology
Amirkabir University of Technology
Tehran, Iran

Behrouz Minaie-Bidgoli
School of Computer Engineering
Iran University of Science and Technology
Tehran, Iran

Mahdi Nasiri
School of Computer Engineering
Iran University of Science and Technology
Tehran, Iran

*Abstract*— Social networks analysis and exploring is important for researchers, sociologists, academics, and various businesses due to their information potential. Because of the large volume, diversity, and the data growth rate in web 2.0, some challenges have been made in these data analysis. Based on definitions, weblogs are a form of social networking. So far, the majority of studies and researches in the field of weblog networks analysis and exploring their stored data have been based on international data sets. In this paper, a framework for preprocessing and data analysis in weblog networks is presented and the results of applying it on a Persian weblog network, as a case study, are expressed.

## I. INTRODUCTION

Social networks have provided a new form of social relationships in cyberspace by using information technology which is expanding with a considerable speed around the world. In 1960, for the first time, a topic as social network was introduced in Illinois University in United States. After that, in 1997, the first social network site launched to address "SixDegrees.com" [16]. According to statistics, number of social networks in the world has become more than a few hundred millions which Facebook with 1.4 billion members is the largest and the most popular social network among them. LinkedIn, Twitter, and Google+ are in the next levels and 58% of internet users are member of at least one social network [17].

The first national social network in Iran was launched in almost ten years ago named "Cloob". There is not any formal statistics about Persian social networks growth, members, features, and activities; however, in 2009, while among the top 20 worldwide popular sites, 8 sites were social networks; in Iran this statistics was 13 social networks sites among the top 20 popular sites, which some of these 13 sites were Persian weblog providers [6]. In all the web mining, data mining, text mining, and social networks analysis studies and researches, data preprocessing is one of the main stages and an integral part of exploring process; however in the most of researches, due to difficulty in obtaining and collecting large volume data, preprocessing stage is done on preprocessed or low volume data to accelerate this step.

According to structure, weblogs are also a manifestation of social networks. Persian weblog networks analysis has been mainly focused on "PersianBlog". These researches have been based on datasets prepared by other researchers and structural data obtained from web crawlers. Also, a low volume data has been processed because of processing limitations [1][10][11]. In another study [3], data stored in a Persian language social network named "ParsiYar" has been studied. This dataset contains information about users' profile, their friendships, and their membership in social groups which its preprocessing results are in [3].

Based on social network analysis definition, weblogs service providers are a form of social network because of their structure. In the simple form, each weblog is node and links between them are considered as edges. Due to special characteristics of Persian language (such as encoding, font, etc.), Persian weblog creation requires special facilities; this is why number of Persian weblogs was low before Persian language specific hosts adventure. Fortunately, at present, there is considerable number of hosts for Persian weblogs, such as Blogfa, ParsiBlog, MihanBlog, and PersianBlog [11]. Persian language people have a great tendency to weblog creation; so that, according to statistics, at present, Iran by 700 thousands blogs is the ninth country based on number of blogs. However, not all of these weblogs are active and number of every few days updated of them is 400 thousands [15].

This paper proposes a framework for weblog networks data preprocessing to social networks exploring and analyzing. The rest of paper is as follow: in section 2, weblogs environment as a social network are studied and some definitions are presented. Section 3, presents proposed framework and section 4, demonstrates the result of applying it on this paper case study. Finally, section 5 concludes the paper.
.

## II. EVALUATION OF WEBLOGS ENVIRONMENT AS SOCIAL NETWORKS

Blog or weblog is a user generated website consists of a series of entries in the form of newspaper arranged in reverse chronological order [18]. Weblogs often contain news or comments about specific topics such as politics, society, and local news. Also, someone can use them for writing content and personal logbooks. A weblog is a combination of text, images, and links to other weblogs. Commenting on content is one of the most important interactive features of weblogs. Kumar and his colleagues have shown that the size of blog environment and communities formed in them have developed with considerable growth rate from 2001 [8].

Among Persian language service provider weblogs, Blogfa, MihanBlog, PersianBlog, and Blogsky, which according to statistics provided by Alexa.com ranked 3, 5, 9, and 25 respectively, are the most famous and the most popular weblogs in Iran.

### A. Weblog's Components

Each weblog service provider can offer unique features to its users. This section introduces the public and common components of weblogs which exist in most of them.

- Post or Entry: a new content which blogger add to her weblog each time.
- Links in posts or Citations: it is possible to link to another post (post to post) or another weblog (post to weblog) or website in a post.
- Blog roll: in list of another blogs that a blogger might recommend due to interest, importance, or similarity by providing links to them (usually in a sidebar list).
- Output link: a link which blogger directly mention it in blog roll.
- Input link: a link through which (usually in blog roll) users enter another blog.
- Comment: a post can be evaluated, praised, and criticized by others. The contents which are based on users' opinion on blog post are comments in the blogosphere.
- Trackback: Trackback is a reverse link through which writer can be aware of links that other have made to her weblog. For this, both weblog service providers must activate trackback system.

### B. Links Structure in Weblogs Social Network

By understanding blogosphere, weblog environment structure is different from general web pages. A blogger has different interactions with another blogger in the field of weblog, such as comments, trackback, and so on. Implicit link information such as post to post or post to blog links are also created in blogosphere. In general, there are four interactions in blogosphere containing comments, trackback, citation, and blog roll, which through them weblog special communicative structure is made.

- Blog roll link: Blog roll is a blog list in the blog main page which contains links to other blogs. Blog rolls have a direct impact on a log popularity rating. A weblog with links to popular weblogs in its blog roll is considered as a well centric weblog. Also, a weblog could be popular if it has links to centric weblogs in its blog roll.
- Citation link: Weblogs consist of many posts by different interests and tendencies. A post allows writer to have a distributed conversation which in it a post can be response to another weblog's post. These references are called "entry to entry" or "post to post" or "citation link".
- Comments link: Bloggers could have a simple and efficient relation with their readers through comments [13]. Commenting systems usually are implemented as an answer set which are arranged chronologically. A blogger by commenting on a weblog post makes a link between this weblog and the destination one. Weblogs by more comments have a higher rate.
- Trackback link: Trackback is a system that allows blogger to know who has seen the blog and commented in it.

### C. Data Types in Social Networks

Using service provided by social networks, micro blogs and weblogs by users lead to creating and saving different data types in data bases. In information related topics, there are three data types: structured, semi structured, and unstructured. So, weblogs network data classification would be as follow:

- Structured data are generated by computing machines and computers and their management, process, and saving is easier that unstructured data. In weblog networks, all system and user interactive information which are stored in relational data bases following table formal structure and associated data models, are in this category, such as each blog links, posts' identification and so on.
- Unstructured data are a form of structured data which do not follow table formal structure and associated data models. However, they have labels and indexes which separate semantic components from each other and create a field and record hierarchical between data [14]. Users' stored information in each blog is in this category.
- Unstructured data are generated by users. More than 90% of world digital data are unstructured which are rapidly growing. Social networks and weblogs are the largest unstructured data creators; these data increase internet traffic, daily [13]. The weblogs' posts, shared images, audios and videos, comments, and so on are the samples of unstructured data.

## III. A FRAMEWORK FOR WEBLOG NETWORKS DATA PREPROCESSING

Data preprocessing is one of the main stages and integral part of analysis and exploring process. The impletion of pre-processing techniques before exploring and analysis of same can improve exploring process lead to significantly executive time reduction. Analysis and exploring methods and algorithms could be applied on proper and structured data [2]. So, two main transformations are needed:

- Raw and semi structured data to structured data
- Raw and unstructured data to structured data

As said before in 2-3, more that 90% of digital data are unstructured; so, the second transformation is time consuming and complicated. By considering blogosphere's information, these data can be categorized in 3 classes which are content data, communication data, and profile data. Due to difference and variation in data, preprocessing stage have all of challenges exists in data mining, text mining, and web mining preprocess. Figure 1 demonstrates the proposed framework for Persian language weblogs' data preprocessing. This framework includes content data preprocessing, communication data preprocessing, and profile data preprocessing as three main parts.

*A. Content Data Preprocessing*

Preprocessing steps applied on content data for computing weblogs similarity are as follow:

- Removing HTML labels
- Content unification: textual data regardless of incorrect spelling, are stored in English, Persian, Persian spoken, and Finglish formats. So, there is some equivalent for each Persian word. Thus, these unstructured data must be unified [4].
- Identify and removing "stop words" [4].
- Extracting keywords
- Creating word vectors
- Computing weblogs similarity

*B. Structure-based Data Preprocessing*

Structure-based data preprocessing steps in order to construct weblogs communication network are as follow:

- Extracting all links: In this step all blogosphere links are extracted includes following steps: Extracting weblogs' blog rolls, extracting comment links between weblogs, and Extracting post links between weblogs.
- Ignoring data set output links: Output links to outside blogs or other sources on the web (images, videos, other web pages) are deleted.

- Deleting inside edges: Internal edges do not show information propagation. So, the link between one node to itself would be deleted.
- Different graphs information combination: The information of three graphs (blog roll, post, and comments) are combined for extracting new information in this step.
- Deleting additional nodes: Nodes with no output link are considered as isolated nodes. Removing these nodes leads to sparseness reduction of blogosphere graph.

*C. Profile Data Preprocessing*

Regardless of data type and its ownership, created and stored information in weblogs network data bases and users' profile can be also categorized in following groups. It should be mentioned that some of these data groups are explicit and some are implicit. So, data mining, text mining, and web mining methods must be used for extracting this information. Regarding weblogs network rules and user settings, some of this information are public and some are private.

- Demographic information, such as age, nationality, gender, education, and so on.
- Information related to products and trademarks, people, places, and so on. This information are available from feedbacks provided by users in comments; it can be also gathered from web page of product owners, people, locations, etc.
- Psychological information; these are features related to people characteristics, values, behavior and attitude, interests, and life style. This information is gathered from advanced user's profile that contains interests, values, and so on. These data can be also obtained by exploring and analyzing user's shared images, videos, etc.
- Behavioral data; Past specific behavior and actions which can show what user is going to do in future. In this manner, the history of linking, commenting, etc. are a basis for user behavior prediction.
- Introductory information; these non-verbal information are shown based on ratings and users' interests to a weblog, post, or comment.
- Positional information; user physical location at present or at any time can be extracted from different blog services.
- User's tendency information; desired products and future planned activities are in this category. This imprecise information can be identified and obtained by prediction methods.

## IV. EXPERIMENTAL RESULT OF PARSIBLOG PREPROCESSING

The case study used in this paper is data stored in a Persian language blog host database named "ParsiBlog" (www.parsiblog.com). This dataset contains some weblogs information which the most important of them are weblogs' archived and not archived posts (post subject, data, text, etc.), each post's comments, blog roll information, users' profile information, blog subjects.

### A. Content -based Data Preprocessing of Data Set

By considering proposed preprocessing framework presented in section 3, content data preprocessing was done on 133472 posts associated to 2149 bloggers written during April to September, 2010. At first, active bloggers' posts were selected. Active bloggers are those who have written at least 6 posts during mentioned time (updated their weblogs at least once per month). 1727 weblogs are known as active blogger which have written 123000 posts during 6 months. The subject and content of these weblogs' posts were preprocessed based on steps described in section 3-1. After doing all preprocessing steps, the number of keywords was about 15000 words. The weblog matrix of previous step words was normal based on TF/IDF [4] criteria and weblogs similarity was computed using Cosine similarity measure [4].

### B. Structure-based Data Preprocessing of Data Set

As said in section 3, there are 4 link types in blogosphere: blog to blog link, post to post link, comment on a post link, and trackback. In ParsiBlog site trackback link is not possible. For example, based on blog to blog link, blogs relations can be modeled as a graph which in it blogs are considered as nodes and their links to each other are directed edges. This mapping is also available through other links. Operations mentioned in section 3-2 were applied on ParsiBlog's data for preparing them to analysis. The features of ParsiBlog's different graphs are illustrated in table 1.

Most of the networks consist of a large number of strongly connected components which in them, there is one connected component with the most nodes [12]. To reduce data sparseness problem, it is possible to only select strongly connected components with large size. In the network of ParsiBlog's blogs with 21305 nodes and 257316 edges, 11706 strongly connected components were identified. It is noteworthy that more than 10000 nodes in this dataset are isolated. The largest connected component has 8933 nodes and 220706 edges. In this paper, strongly connected component with at least 10 nodes were selected. So, the final graph at this stage is composed of 9065 nodes and 222216 edges. Figure 3, demonstrates strongly connected components distribution in ParsiBlog network. The preprocessed weblogs network is illustrated in figure 2. Weblogs network information comparing before and after preprocessing is given in table 2.

To obtain weblogs popularity, they were rated based on PageRank [5] and HITS [7] ranking algorithms and also number of input links. One of the simplest methods for ranking weblogs is using their input links which through it weblog by more input links has more popularity (Figure 5).

In HITS algorithms there are two different weblogs named hub and authority [7]. Authority blog contains important contents and hub blog, such as reference list, is used for direct users to other authority blogs. Thus, a good hub blog should have links to a large number of good authority blogs in the same field. A blog can be a good hub and also a good authority, at once, such as AliShariaty blog (www.Alishariaty.parsiblog.com). A recursive algorithm is implemented and at first, the hub and authority values are set to 1. The algorithm would be converged after 60 iterations. PageRank algorithm was also applied on the graph by using .85 as damping factor. This means that each page ranks the pages it links to by a value less that itself. The more output link lead to less rating for linked pages. This algorithm is repeated recursively until rates adjusted and not changed [5].

### C. Profile-based Data Preprocessing of Data Set

Statistic information presented in this section are based on bloggers' information preprocessing. Bloggers' statistic and demographic information are available through their profile information. Bloggers average age is about 21 years. Statistic shows that most of users are young in the range of 15 to 30 years. Also, bloggers gender studies show that most of bloggers are men; even, number of male bloggers is twice the number of females. Most of bloggers are single with diploma to bachelor education; also, bachelor frequency is more than other educational level.

Based on analysis done on posts writing time, it was cleared that in early hours of day, between 1 to 6 AM, writing posts is decreased because at this time most of bloggers are asleep. The minimum of diagram is also between 3 to 5 AM. From 16 PM to before midnight, the number of posts is increased, because at this time bloggers are free and like to update their weblogs (Figure 4).

The frequency of April's post is the most which is about 5 or 6 percent more than other months. This is because of Norouz holidays which in them bloggers are free. Total number of comments in the first six months of 2010 is equal to 119280 which are related to 25408 posts (however, total number of posts is equal to 133471). The average number of comments for each post equals to 0.89; the average number less that 1 shows that there are many posts with no comment. Usually, only the posts of popular weblogs have comments. There are only about 1866 posts with more than 10 comments; more than 90% of weblogs' posts either have no comment or have less than 5 comments.

## V. CONCLUSION

Weblogs as a manifestation of network and social structure are a representation of communities in real world from different aspect such as cultural, social, religious, political, etc. Different countries and nations, regardless of having common human and social aspects, because of differences in religion, culture, education, law, etc., have different behavioral patterns, values, interaction, etc. So, it is not correct to use foreign and international samples results for internal ones and deciding based on the.

In this paper a framework was presented for weblog networks data preprocessing and its results were implemented on a Persian language weblog host named ParsiBlog. No other study in the field of Persian language weblog data preprocessing was done till now. This paper's data were analyzed and explored at their first time; so, the study's main data preparation and construction were done by this paper authors, despite of other academic researches. ParsiBlog preprocessed dataset was used by authors for presenting a recommender system in social networks [1].

According to weblogs and social networks construction similarity, their importance and growth, and also their transformation to a social media in web environment, their efficiency, and national conditions, weblogs social network identification and analysis is useful for growth and survival of a weblog service provider. Content analysis of weblogs' posts and their classification is considered as future work. Also, based on its result and users' profile analysis, tendency of users with different characteristics to writing different posts would be studied. Link analysis can be also used for specifying users' tendency to different subjects.

TABLE I. COMPARING DIFFERENT NETWORKS' FEATURE IN PARSIBLOG (CC = CONNECTED COMPONENT)

| Network | # Nodes | # Edges | Degree avg. | Density | # Strongly CC |
|---|---|---|---|---|---|
| *Weblogs net* | 21305 | 257316 | 24.15 | 0.000567 | 11706 |
| *Comments net* | 11187 | 92703 | 16.57 | 0.000741 | 8215 |
| *Posts net* | 4664 | 10528 | 4.51 | 0.000484 | 4146 |

TABLE II. . WEBLOGS NETWORK INFORMATION COMPARING BEFORE AND AFTER PREPROCESSING

| Network | # Nodes | # Edges | Degree avg. | Density | Clustering Coefficient |
|---|---|---|---|---|---|
| *Primary net* | 21305 | 257316 | 24.1554 | 0.000567 | 0.31747 |
| *Preprocessed net* | 9065 | 222216 | 49.027248 | 0.002704 | 0.37995 |

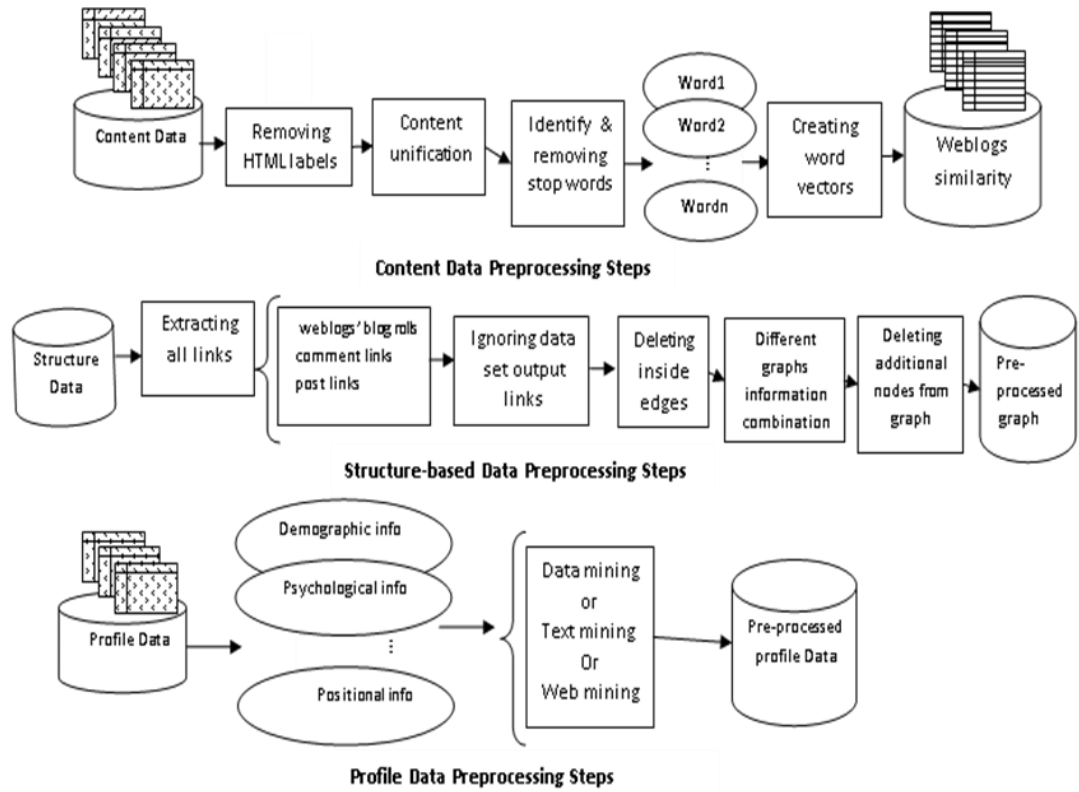

Figure 1.  The Framework for Weblog Networks Data Preprocessing

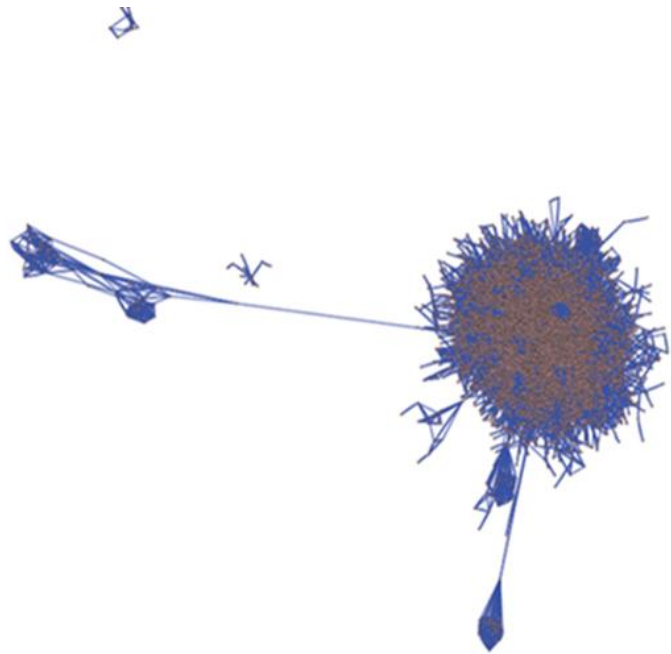

Figure 2.  Data preprocessed social network of ParsiBlog

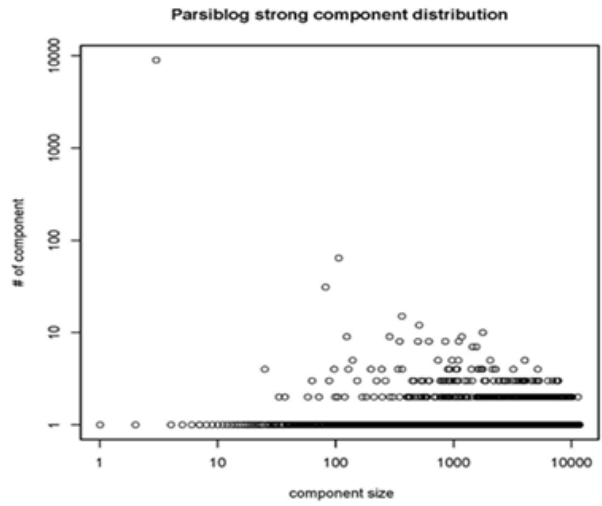

Figure 3. Strongly connected components distribution in ParsiBlog

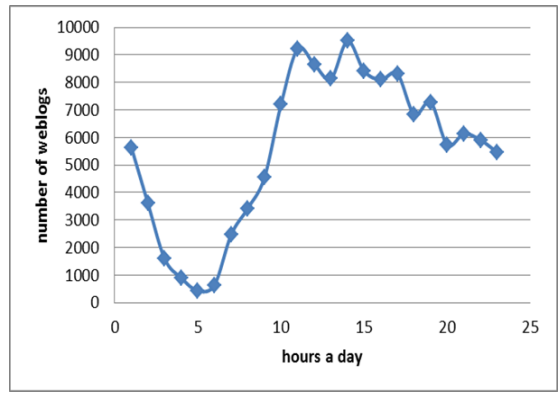

Figure 4. Blogs frequency based on Hours a day

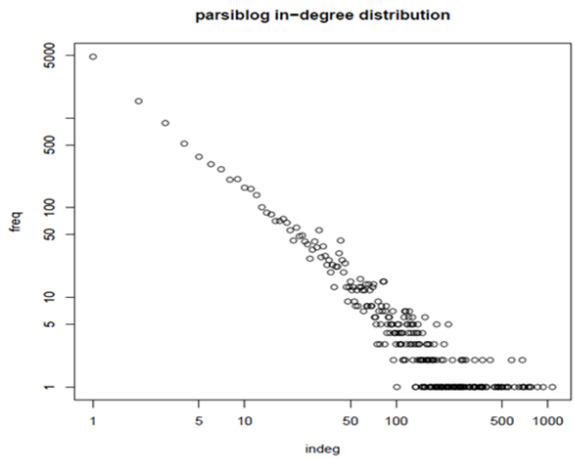

Figure 5. Blogs frequency based on input links